\documentclass[prb,twocolumn,superscriptaddress,amsmath,amssymb]{revtex4}

\usepackage{array}
\usepackage{graphicx} 
\usepackage{color}
\usepackage{bm}
\usepackage{pstricks}

\def\beq{\begin{equation}}
\def\eeq{\end{equation}}
\def\bea{\begin{eqnarray}}
\def\eea{\end{eqnarray}}
\def\fatR{{\bf R}}

\def\fatM{{\bf M}}

\begin{document}
\title{Machine Learning of Molecular Electronic Properties in Chemical Compound Space} 

\author{Gr\'egoire Montavon}
\affiliation{Machine Learning Group, Technical University of Berlin, Franklinstr 28/29, 10587 Berlin, Germany}
\author{Matthias Rupp}
\affiliation{Institute of Pharmaceutical Sciences, ETH Zurich, 8093 Z{\"u}rich, Switzerland}
\author{Vivekanand Gobre}
\affiliation{Fritz-Haber-Institut der Max-Planck-Gesellschaft, 14195 Berlin, Germany}
\author{Alvaro Vazquez-Mayagoitia}
\affiliation{Argonne Leadership Computing Facility, Argonne National Laboratory, Argonne, Illinois 60439, USA}
\author{Katja Hansen}
\affiliation{Fritz-Haber-Institut der Max-Planck-Gesellschaft, 14195 Berlin, Germany}
\author{Alexandre Tkatchenko}
\email{tkatchen@fhi-berlin.mpg.de}
\affiliation{Fritz-Haber-Institut der Max-Planck-Gesellschaft, 14195 Berlin, Germany}
\affiliation{Department of Chemistry, Pohang University of Science and Technology, Pohang 790--784, Korea}
\author{Klaus-Robert M\"uller}
\email{klaus-robert.mueller@tu-berlin.de}
\affiliation{Machine Learning Group, Technical University of Berlin, Franklinstr 28/29, 10587 Berlin, Germany}
\affiliation{Department of Brain and Cognitive Engineering, Korea University, Anam-dong, Seongbuk-gu, Seoul 136-713, Korea}
\author{O. Anatole von Lilienfeld}
\email{anatole@alcf.anl.gov}
\affiliation{Argonne Leadership Computing Facility, Argonne National Laboratory, Argonne, Illinois 60439, USA}
\date{\today}

\begin{abstract}
The combination of modern scientific computing with electronic structure theory can lead to an unprecedented 
amount of data amenable to intelligent data analysis for the identification of meaningful, novel, and predictive structure-property relationships. 
Such relationships enable high-throughput screening for relevant properties
in an exponentially growing pool of virtual compounds that are synthetically accessible. 
Here, we present a machine learning (ML) model, trained on a data base of \textit{ab initio} calculation results for thousands of organic molecules,
that simultaneously predicts multiple electronic ground- and excited-state properties.
The properties include atomization energy, polarizability,
frontier orbital eigenvalues, ionization potential, electron affinity, and excitation energies. 
The ML model is based on a deep multi-task artificial neural network, 
exploiting underlying correlations between various molecular properties. 
The input is identical to \emph{ab initio} methods,
\emph{i.e.}~nuclear charges and Cartesian coordinates of all atoms.
For small organic molecules the accuracy of such a ``Quantum Machine'' 
is similar, and sometimes superior, to modern quantum-chemical methods---at negligible computational cost.
\end{abstract}

\maketitle

\section{Introduction}
The societal need for novel computational tools and data treatment that serve the 
accelerated discovery of improved and novel materials has gained considerable momentum in the form of the
materials genome initiative~\cite{WhiteHouseMGI}. 
Modern electronic structure theory and compute hardware have progressed to the point where 
electronic properties of virtual compounds can routinely be calculated with satisfying accuracy.
For example, using quantum chemistry and distributed computing, members of the widely advertised Harvard Clean Energy Project 
endeavor to calculate relevant electronic properties for millions of chromophores~\cite{MGI2011}.
A more fundamental challenge persists, however: 
It is not obvious how to distill from the resulting data the crucial insights that relate structure to property in a predictive and quantitative manner. 
How are we to systematically construct robust models of electronic structure properties that
properly reflect the information already obtained for thousands to millions of different chemical compounds?

With increasing amounts of data and available computational resources, 
increasingly sophisticated statistical data analysis, or machine learning (ML) methods, 
have already been applied to predicting not only outcomes of experimental measurements but also outcomes of computationally 
demanding high-level electronic structure calculations.  
In close analogy to the quantitative structure property relationships (QSPRs) prevalent in
cheminformatics and bioinformatics, QSPRs can also be constructed for electronic structure properties.
Examples include QSPRs for exchange-correlation potentials using neural networks (NNs)~\cite{NN4B3LYP_Chen2003,NN4B3LYP_Chen2004},
basis-set effects using support vector machines~\cite{Balabin09,SVM4CBS_Lomakina2011}, or 
molecular reorganization energies affecting charge transfer rates~\cite{RatnerJACS2005,anatole-MilindDenis2011},
or for solid ternary oxides~\cite{MachineLearningHautierCeder2010}.
Ordinarily, these applications rely on association, using regression methods that create statistically optimized relationships between
so called descriptor variables and electronic property of interest. 
Not surprisingly, the heuristic {\em ad hoc} identification and formatting of appropriate descriptors 
represents a crucial and challenging aspect of any QSPR, and is to be repeated for every property and class of chemicals.  

We make use of an alternative ML approach, recently introduced by some of us
for the modeling of molecular atomization energies~\cite{RuppPRL2012}.
This approach is based on a strict first principles view on chemical compound space~\cite{CCSanatole2013}.
Specifically, solutions to Schr\"odinger's equation (SE) are inferred for organic query molecules using 
the same variables that also enter the electronic Hamiltonian $H$, 
i.e.~nuclear charges $Z_I$ and positions $\fatR_I$,~\footnote{The number of electrons is implicitly encoded by imposing charge neutrality.}
 and that are mapped to the corresponding total potential energy, 
$H(\{Z_I, \fatR_I\}) \stackrel{\rm \Psi}{\longmapsto} E$.~\cite{HK,CCSanatole2013}
Unlike the aforementioned QSPRs this ML model is free of any heuristics: It exactly encodes the supervised learning problem posed by SE, 
i.e.~instead of finding the wavefunction $\Psi$ which maps the system's Hamiltonian to its energy, 
it directly maps system to energy (based on examples given for training), $\{Z_I, \fatR_I \} \stackrel{\rm ML}{\longmapsto} E$.
The employed descriptor, dubbed ``Coulomb''-matrix, is directly obtained from $\{Z_i,\fatR_I\}$.
As such this constitutes a well defined supervised-learning problem and in the limit of converged number of
training examples the ML model becomes a formally exact inductive equivalent to the deductive solution of SE. 
It is advantageous that the training data can come from experiment just as well as from numerical evaluation 
of the corresponding quantum mechanical observable using approximate wave-functions (separated nuclear and 
electronic wavefunctions, Slater determinant expansions etc.), Hamiltonians (such as H\"uckel or any exchange-correlation potential), 
and self-consistent field procedures. 

\begin{figure*}[htbp]
\centering
\includegraphics[trim=2mm 5mm 0 0,scale=0.38, angle=0]{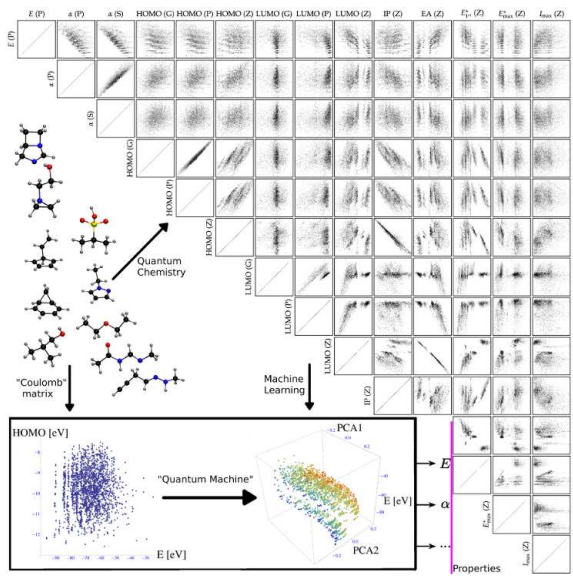}
\caption{
Overview of calculated data base used for training and testing of the Machine Learning model. 
Quantum chemistry results for 14 properties of 7211 molecules are displayed. All properties and level theory, GW (G), PBE0 (P), and ZINDO (Z), are defined in Section \ref{section-properties}.
Cartoons of 10 exemplary molecules from the data base are shown, 
they are used as input for quantum chemistry, for learning, or for prediction.  
Relying on input in ``Coulomb'' matrix form, 
the concept of a ``Quantum Machine'' is illustrated for two seemingly uncorrelated properties, 
atomization energy $E$ and HOMO eigenvalue, which are decoded in terms of the two largest principal components (PCA1,PCA2) 
of the last neural network layer for 2k molecules, not part of training. 
The color-coding corresponds to the HOMO eigenvalues.
}
\label{fig:Correlation}
\end{figure*}

Building on our previously introduced work~\cite{RuppPRL2012}, we here present a more mature ML model 
developed to accomplish the following two additional tasks, 
(i) simultaneously predict a variety of different electronic properties for a single query, and
(ii) reach an accuracy comparable with the employed reference method used for generating the training set. 
The presented ML model is based on a {\em multi-task} deep artificial NN approach that captures correlations between 
seemingly related and unrelated properties and levels of theory. 
Remarkable predictive accuracy for ``out-of-sample'' molecules ({\em i.e.}~molecules that were not part of the training set) 
has been obtained through the use of random Coulomb-matrices that introduce invariance with respect to atom indexing.
For training  we generated a quantum chemical data base containing nearly 10$^5$ entries for over seven thousand stable organic molecules, 
made of up to 7 atoms from main-group elements, consisting of C, N, O, S, and Cl, saturated with hydrogens to satisfy valence 
rules.~\cite{ReymondChemicalUniverse, ReymondChemicalUniverse2}
For each molecule atomization energy, static polarizabilities, frontier orbital eigenvalues, and excitation energies and intensities
have been calculated using a variety of widely used electronic structure methods, including state-of-the-art
first principles methods, such as hybrid density-functional theory and the many-body {\em GW} approach (see Methods and Ref.~\cite{DetailsDataNote}). 
Fig.~\ref{fig:Correlation} illustrates the complete property data base, and how it has been used for model training and prediction.

\section{Methods}

\subsection{Molecular structures (input)}

While the present ML model approach is generally applicable, for the purpose of this study
we restrict ourselves to the chemical space of small organic molecules.
For all the cross-validated training and out-of-sample model performance testing, 
we rely on a controlled test-bed of molecules,
namely a subset of the GDB-13 data base~\cite{ReymondChemicalUniverse, ReymondChemicalUniverse2}
consisting of all the 7211 small organic molecules that have up to 7 second and third row atoms
consisting of C, N, O, S, or Cl, saturated with hydrogens.
The entire GDB-13 data base represents an exhaustive list of the $\sim$0.97B organic
molecules that can be constructed from up to 13 such ``heavy'' atoms.
All GDB molecules are stable and synthetically accessible according to organic chemistry rules~\cite{ReymondChemicalUniverse3}.
Molecular features such as functional groups or signatures include single, double
and triple bonds; (hetero-)cycles, carboxy, cyanide, amide, amines, alcohol, epoxy, sulfide,
ether, ester, chloride, aliphatic, and aromatic groups.
For each of the many possible stoichiometries, many constitutional isomers are considered, each being
represented only by a single conformational isomer.

Based on the string representation (SMILES \cite{w1988,www1989}) of molecules in the data base, 
we used the universal force-field~\cite{UFFRappeGoddard1992} to generate reasonable Cartesian molecular
geometries, as implemented in {\tt OpenBabel}~\cite{OpenBabel}.
The resulting geometries were relaxed using the PBE approximation~\cite{PBE} to Kohn-Sham DFT~\cite{KS} 
in converged numerical basis, as implemented in the {\tt FHI-aims} code~\cite{aims} (tight settings/tier2 basis set).
All geometries are provided in the supplementary material.

\subsection{Molecular representation (descriptor)}
One of the most important aspects for creating a functional ML model is the choice of an appropriate data representation
(descriptor) that reflects important constraints and properties due to the underlying physics, SE in our case. 
While there is a wide variety of descriptors used in chem- and bio-informatics 
applications~\cite{SchneiderReview2010,SignatureFaulon2003,WienerDescriptors,TodeschiniConsonniHandbookDescriptor,DescriptroOverviewMeringer2005}
they conventionally are based on prior knowledge about chemical binding, electronic configuration, 
or other quantum mechanical observables.  
Instead, we derive our representation without any pre-conceived knowledge, 
i.e.~exclusively from stoichiometry and configurational information, 
from the generated according to the previous subsection.
As such, the molecular representation is in complete analogy to the electronic Hamiltonian used in {\em ab initio} methods.

For this study we use a randomized variant of the recently introduced ``Coulomb matrix'', $\fatM$.~\cite{RuppPRL2012}.
The Coulomb matrix is an inverse atom-distance matrix representation 
that is unique ({\em i.e.}~no two molecules will have the same Coulomb matrix unless they are identical or enantiomers), 
and retains invariance with respect to molecular translation and rotation by construction.
\bea
M_{IJ}  =
\begin{cases}
  0.5 Z_I^{2.4} & \mbox{for} \;\; I = J,\\
   \frac{Z_IZ_J}{|\fatR_I - \fatR_J|} & \mbox{for} \;\; I \ne J.
\end{cases}
\label{eq:matrix}
\eea
Off-diagonal elements encode the Coulomb repulsion between nuclear charges of atoms $I$ and $J$,
while diagonal elements represent the stoichiometry through an exponential fit in $Z$ to the free atoms' potential energy.
We have enforced invariance with respect to atom indexing by representing each molecule by a probability distribution over Coulomb matrices $p(\fatM)$ generated by different atoms indexings of the same molecule. Details for producing such random Coulomb matrices are given in the supplementary material.

\subsection{Molecular electronic properties (output)}
\label{section-properties}
The reference values necessary for learning and testing consist of various electronic ground and excited-state
properties of molecules in their PBE geometry minimum.
Specifically, we consider atomization energies $E$, static polarizabilities (trace of tensor) $\alpha$, 
frontier orbital eigenvalues HOMO and LUMO, ionization potential IP, and electron affinity EA. Furthermore, from optical spectrum simulations (10nm-700nm), we consider first excitation energy $E^*_{1^{st}}$, excitation of maximal optimal absorption $E^*_{max}$, and its corresponding intensity $I_{max}$.
Data ranges of properties for the molecular structures and for various levels of theory are given in footnote~\cite{DetailsDataNote},
property mean-values in the data set also feature in Table~\ref{tab:error}.

To also gauge the impact of the reference method's level of theory on the ML model, 
polarizabilities and frontier orbital eigenvalues were evaluated with more than one method.
Static polarizability has been calculated using self-consistent screening (SCS)~\cite{MBD} as well as
hybrid density functional theory (PBE0)~\cite{PBE0,PBE01}.
PBE0 has also been used to calculate atomization energies and frontier orbital eigenvalues. 
Electron affinity, ionization potential, excitation energies, and maximal absorption intensity
have been obtained from Zerner's intermediate neglect of differential overlap (ZINDO)\cite{zindo1,zindo2,zindo3}.
Hedin's GW approximation~\cite{GW} has also been used to evaluate frontier orbital eigenvalues.
GW is a quasi-particle ab initio many-body perturbation theory, known to accurately
account for electronic excitations that describe electron addition and removal processes~\cite{GW}.
The SCS, PBE0, and GW calculations have been carried out using {\tt FHI-aims}~\cite{aims,aimsgw},
ZINDO/s calculations are based on the {\tt ORCA} code~\cite{orca}.
ZINDO/s is an extension of the INDO/s semiempirical method with
parameters to accurately reproduce  single excitation spectra of
organic compounds and complexes with rare earth elements. The INDO
Hamiltonian neglects some two-center two-electrons integrals in order
to simplify the calculation of time-dependent Hartree-Fock equations.
While ZINDO results are usually not as accurate as highly correlated
methodologies the semiempirical Hamiltonian reproduces the most important features
of the absorption spectra of many small molecules and complexes,
particularly characterizing their most intense bands on the UV-Vis spectra.
All properties are provided in the supplementary material.

Similar conclusions hold for the selected levels of theory: The employed methods can be considered
to represent a reasonable compromise between computational cost and predictive accuracy. 
It should be mentioned, that ML methods can, in principle, be applied to any method or level of approximation.

\subsection{Training the model}
Our model consists of a deep and multi-task neural network~\cite{Caruana97,Bengio07}  that is trained on molecule-properties pairs. 
It learns to map Coulomb matrices
to all the 14 properties of the corresponding molecule simultaneously.
NNs are well-established for learning functional relationships between input and output. 
They have successfully been applied to varying tasks such as object recognition\cite{haffner98} and speech recognition\cite{waibel}.
Given a sufficiently large NN, its universal approximation capabilities~\cite{cybenko}, 
and the existence of the underlying noise-free Schr\"odinger equation, 
a NN solution can be expected to exist that satisfyingly relates molecules to their properties. 
Specifically, a deep NN will properly unfold, layer after layer, a complex input 
into a simple representation of molecular properties. 
Finding the true relationship unfolding among those that fit the 
training data can be challenging because there is typically a manifold of solutions. 
The multi-task set up forces the NN to predict multiple properties simultaneously. 
This is conceptually appealing because these additional constraints narrow down the search for
the ``true model'' \cite{baxter} as the set of models that fit all properties simultaneously is smaller. Details about the neural network training procedure are provided in the supplementary material.

\section{Results and discussion}
Before reporting and discussing our results, we note the long history of statistical 
learning of the potential energy hyper surface for molecular dynamics applications. 
It includes, for example, the modeling of potential energies surfaces with artificial neural networks 
starting with the work of Sumpter and Noid in 1992~\cite{Sumpter92,Neuralnetworks_Scheffler2004,Neuralnetworks_BehlerParrinello2007,MachineLearningWaterPotential_Handley2008,Handley10,Neuralnetworks_Behler2011}, 
or Gaussian processes \cite{bpkc2010,Mills11}. 
Our work aims to move beyond single molecular systems and learn to generalize to unseen compounds. 
This extension is not trivial, as the input representation must deal with molecules 
of diverse sizes and compositions in the absence of one-to-one mapping between atoms of different molecules.

\subsection{Database}
Scatter plots among all properties for all molecules are shown in Fig.~\ref{fig:Correlation}. 
Visual inspection confirms expected relationships between various properties: 
Koopman's theorem relating ionization potential to the HOMO eigenvalue~\cite{ChemistsGuidetoDFT},
hard soft acid base principle linking polarizability to stability~\cite{HSABPearson},
or electron affinity correlating with first excitation energy.
Correlations of identical properties at different levels of theory reveal more subtle differences.
Polarizabilities, calculated using PBE0, or with the more approximate self-consistent screening (SCS) model~\cite{MBD}, are strongly correlated.
Also less well known relationships can be extracted from this data. 
One can obtain to a very decent degree, for example, the GW HOMO eigenvalues by subtracting 1.5 eV from the corresponding PBE0 HOMO values. 

Some properties, such as atomization and HOMO energies, exhibit very little correlation in their scatter plot.
The inset of Fig.~\ref{fig:Correlation} illustrates how our Quantum Machine (i.e. NN based ML model) extracts and exploits hidden 
correlations for these properties despite the fact that they can not be recognized by visual inspection.
Similar conclusions hold for atomization energy versus first excitation energy, or polarizability versus HOMO. 

\subsection{Accuracy vs.~training set size}
\begin{figure}[htbp]
\centering
\includegraphics[scale=0.55, angle=0]{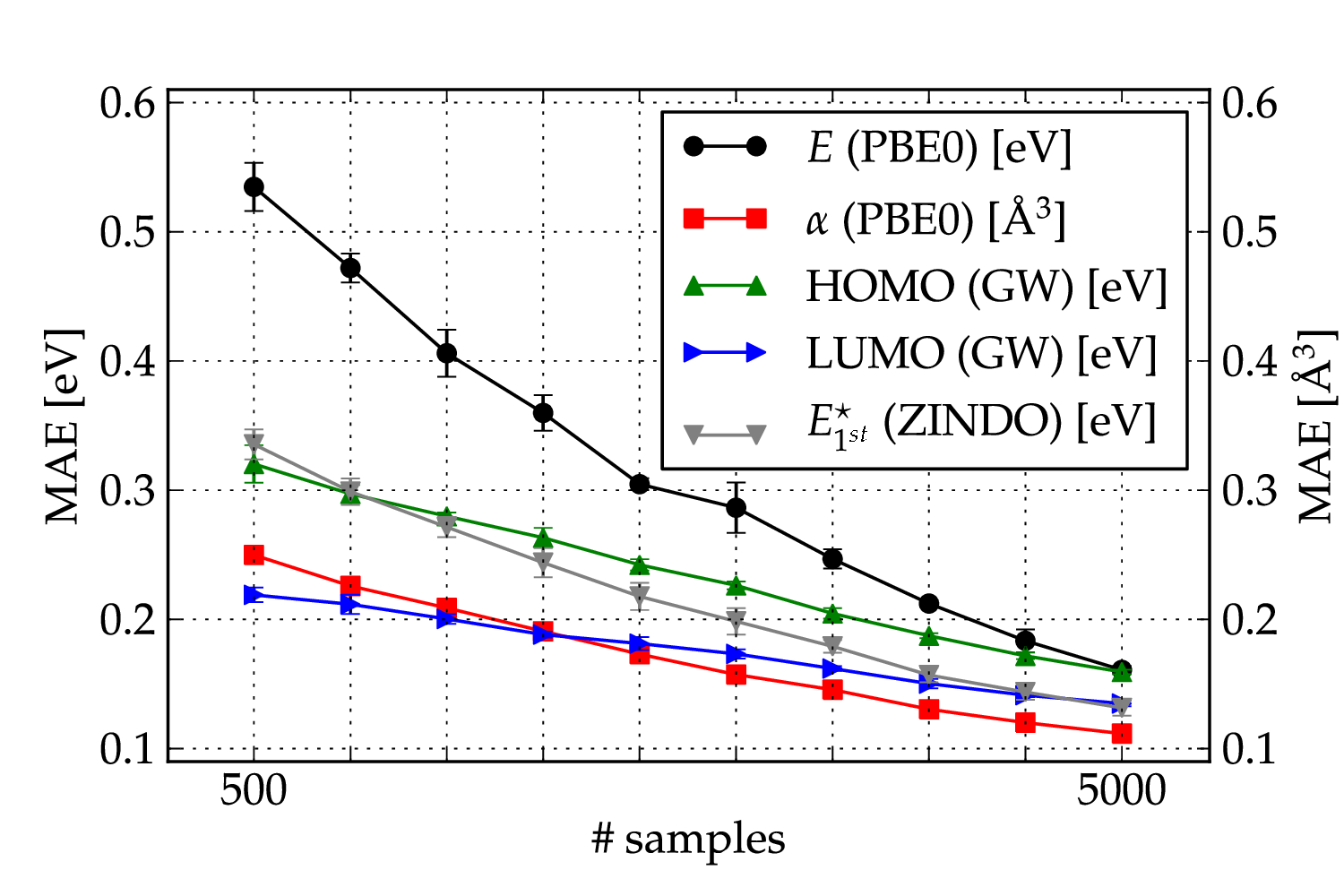}
\caption{
Error decay of ML model with increasing number of molecules in training set (shown on a logarithmic scale). 
The MAE and its error bar is shown for atomization energy ($E$), 
polarizability ($\alpha$), frontier orbital eigenvalues (HOMO, LUMO), 
and first excitation energy ($E^*_{1^{st}}$).
}
\label{fig:Saturation}
\end{figure}

It is an important feature of any ML model that the error can be controlled systematically as the training size is varied. 
We have investigated this dependence for our ML model.
Fig.~\ref{fig:Saturation} shows a typical decay of the ML model's mean absolute error (MAE) 
for predicting properties of ``out-of-sample'' molecules as the number of molecules in training set increases
logarithmically from 500 to 5000, its maximal value in the data base of 7211. 
For all investigated properties, the improvement of error suggests that the MAE could still be lowered 
even further through addition of more molecules. 
However, since the reference method's ``precision'' (i.e.~estimated accuracy of the employed level of theory) is reached for almost all properties already using 5000 examples, adding further examples does not make sense. 
For the atomization energy the decay is particularly dramatic: 
A ten-fold increase in number of molecules (500 $\rightarrow$ 5000) reduces the error by 70\%,
from 0.55 to 0.16 eV. But also for the HOMO/LUMO eigenvalues, the error reduces substantially. 
We find that the expected error decay law of $\propto 1/\sqrt{N}$ is only recovered for the atomization energy, 
for other properties the error decays more slowly.
Fig.~\ref{fig:Saturation} also features the statistical error bars for the MAEs---a measure of outliers. 
The error bar is only slightly larger than symbol size, and hardly varies as the training set increases 
and the testing set decreases.

\subsection{Final ML model}
\begin{figure*}[htbp]
\centering
\includegraphics[trim=1cm 5mm 0 5mm,scale=0.49, angle=0]{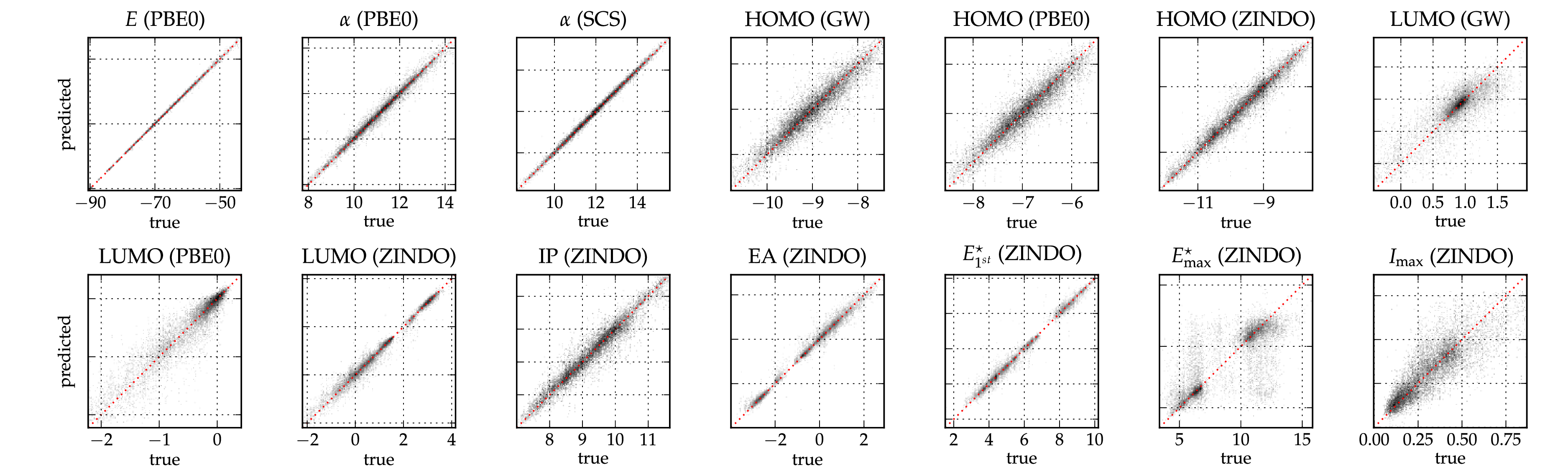}
\caption{
Scatter plot of true value versus ML model value for all properties. The red line indicates the identity mapping.
All units correspond to the entries shown in Table~\ref{tab:error}, 
}
\label{fig:Scatter}
\end{figure*}

After cross-validated training on the largest training set with randomly selected 5000 molecules, 
2211 predictions have been made for the remaining
``out-of-sample'' molecules, yielding at once all the 14 quantum chemical properties per molecule. 
The corresponding true versus predicted scatter plots feature in Fig.~\ref{fig:Scatter}.
The corresponding mean absolute and root-mean square errors are shown in Table~\ref{tab:error}, 
together with literature estimates of errors typical for the corresponding level of theory. 
Errors of all properties range in the single digit \% of the mean property. 
Remarkably, when compared to published typical errors for the corresponding level of theory,
{\em i.e.}~used as reference method for training, similar accuracy is obtained---the 
sole exception being the most intense absorption and its associated excitation energy. 
This, however, is not too surprising: Extracting the information about a particular excitation
energy and the associated absorption intensity requires sorting the entire optical spectrum---thus encoding significant knowledge 
that was entirely absent from the information employed for training.
For all other properties, however, our results suggest that the presented ML model 
makes ``out-of-sample'' predictions with an accuracy competitive to the employed reference methods. 
These methods include some of the more costly state-of-the-art electronic structure calculations, 
such as GW results for HOMO/LUMO eigenvalues and hybrid DFT calculations for atomization energies and polarizabilities. 
Work is in progress to extend our ML approach to other properties, such as the prediction of ionic forces or 
the full optical spectrum.
We note, however, that for the purpose of this study any level of theory and any
set of geometries could have been used.

\begin{table}
\caption{
Mean absolute errors (MAE) and root mean square errors (RMSE) for out-of-sample predictions by ML model, 
together with typical error estimates of corresponding reference level of theory.
Errors are reported for all 14 molecular properties, and are based on out-of-sample predictions for 2211
molecules using a multi-task multi-layered NN ML model obtained by cross-validated training on 5000 molecules. 
The corresponding true versus predicted scatter plots feature in Fig.~\ref{fig:Scatter}.
Property labels refer to level of theory and molecular property, {\em i.e.}~atomization energy ($E^{\rm ref}$), 
averaged molecular polarizability ($\alpha$), 
HOMO and LUMO eigenvalues,
ionization potential (IP), 
electron affinity (EA),
1$^{\rm st}$ excitation energy ($E^*_{1^{st}}$), 
excitation frequency of maximal absorption ($E^*_{\rm max}$), 
and corresponding maximal absorption intensity ($I_{\rm max}$). 
To guide the reader, the mean value of the property across all the 7211 molecules in the data base is shown in the second column.
Energies, polarizabilities, and intensity are in eV, \AA$^3$, and arbitrary units, respectively.
}
\label{tab:error}
\begin{tabular}{|l|c|c|c|c|} \hline
Property                         & Mean    & MAE  & RMSE & Reference~MAE\\ \hline
$E$ (PBE0)                       & -67.79  &  0.16 & 0.36& 
\scriptsize 0.15\footnote{PBE0, MAE of formation enthalpy for G3/99 set~\cite{ScuseriaPerdewJCP2003,CurtissPopleJCP2000}},   
0.23\footnote{PBE0, MAE of atomization energy for 6 small molecules~\cite{TruhlarPCCP2004,TruhlarJPCA2003}}, 
0.09-0.22\footnote{B3LYP, MAE of atomization energy from various studies~\cite{ChemistsGuidetoDFT}}  \\ 
$\alpha$ (PBE0)                  & 11.11   &  0.11 & 0.18& 
\scriptsize 0.05-0.27\footnote{B3LYP, MAE from various studies~\cite{ChemistsGuidetoDFT}},
0.04-0.14\footnote{MP2, MAE from various studies~\cite{ChemistsGuidetoDFT}}\\ 
$\alpha$ (SCS)                   & 11.87   &  0.08 & 0.12& 
\scriptsize 0.05-0.27$^d$,
0.04-0.14$^e$\\
HOMO (GW)                      &-9.09    &  0.16 & 0.22& -\\ 
HOMO (PBE0)                      &-7.01    &  0.15 & 0.21& 2.08\footnote{MAE from GW values}\\ 
HOMO (ZINDO)                     & -9.81   &  0.15 & 0.22& 0.79$^h$\\ 
LUMO (GW)                        & 0.78    &  0.13 & 0.21& - \\ 
LUMO (PBE0)                      &-0.52    &  0.12 & 0.20& 1.30$^h$\\ 
LUMO (ZINDO)                     & 1.05    &  0.11 & 0.18& 0.93$^h$\\ 
IP (ZINDO)                       & 9.27    &  0.17 & 0.26&
0.20\footnote{PBE0, MAE for G3/99 set~\cite{ScuseriaPerdewJCP2003,CurtissPopleJCP2000}},
0.15$^d$\\
EA (ZINDO)                       & 0.55    &  0.11 & 0.18&
0.16$^g$,
0.11$^d$\\
$E^*_{1^{st}}$ (ZINDO)           & 5.58    &  0.13 & 0.31& 
0.18\footnote{ZINDO, MAE for set of 17 retinal analogs~\cite{deLeraJCC2006}},
0.21\footnote{TD-DFT(PBE0), MAE for set of 17 retinal analogs~\cite{deLeraJCC2006}}\\ 
$E^*_{\rm max}$ (ZINDO)          & 8.82    &  1.06 & 1.76& -\\ 
$I_{\rm max}$ (ZINDO)            & 0.33    &  0.07 & 0.12& -\\ 
\hline
\end{tabular}
\end{table}

\begin{figure}
\centering
\includegraphics[scale=0.55, angle=0]{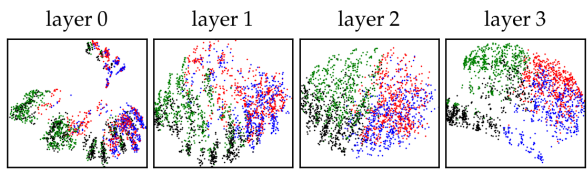}
\definecolor{darkgreen}{rgb}{0,0.5,0}
\caption{
Principal component analysis (PCA) on the multiple layers of the deep neural network. Each point (molecule) is colored according to the rule: $E$ and HOMO large $\rightarrow$ red ; $E$ large and HOMO small $\rightarrow$ blue ; $E$ small and HOMO large $\rightarrow$ green ; $E$ and HOMO small $\rightarrow$ black. We can observe that the neural network extracts, layer after layer, a representation of the chemical space, that better captures the multiple properties of the molecule.
}
\label{fig:MTL}
\end{figure}

The remarkable predictive power of the ML model can be rationalized by 
(i) the deep layered nature of the NN model that permits to progressively extract 
the relevant problem subspace from the input representation and gain predictive accuracy~\cite{Braun2008,Montavon2011}; 
(ii) inclusion of random Coulomb matrices for training, 
effectively imposing invariance of property with respect to atoms indexing, clearly benefits the model's accuracy:
Additional tests suggest that using random, instead of sorted or diagonalized~\cite{RuppPRL2012}
Coulomb matrices, also improves the accuracy of Kernel Ridge Regression models to similar degrees;
and (iii) the multi-task nature of the NN accounts for strong and weak correlations between seemingly unrelated 
properties and different levels of theory. Aspects (i) and (iii) are also illustrated in Fig.~\ref{fig:MTL}.

We reiterate that evaluation of all the 14 properties at said level of accuracy for an out-of-sample
molecule requires only milli-seconds using the ML model, as opposed to several CPU hours using 
the reference methods used for training. The down-side of such accuracy, of course, are the limits in transferability.
All ML model predictions are strictly limited to out-of-sample molecules that interpolate. 
More specifically, the 5000 training molecules must resemble the query molecule in a similar fashion as 
they resemble the 2211 test molecules. For compounds that bear no resemblance to the training set, 
the ML model must not be expected to yield accurate predictions.
This limited transferability might one day become moot through more intelligent choice and construction of 
molecular training sets tailored cover {\em all} of a pre-defined chemical compound space, 
i.e. all of the relevant geometries and elemental compositions, up to a certain number of atoms.

\section{Conclusion}
We have introduced a machine learning (ML) model for predicting electronic 
properties of molecules based on training deep multi-task artificial NNs in chemical space.
Advantages of such a ``Quantum Machine'' (QM) (conceptually speaking, as illustrated in Fig.~\ref{fig:Correlation}) are the following,
(a) multiple dimensions: A single QM execution simultaneously yields multiple properties at multiple levels of theory;
(b) systematic reduction of error: By increasing the training set size the QM's accuracy can be converged to a degree that 
outperforms modern quantum chemistry methods, hybrid density-functional theory and the GW method in particular; 
(c) dramatic reduction in computational cost: The QM makes virtually instantaneous property predictions;
(d) user-friendly character: Training and use of the QM does not require knowledge about electronic structure, or even about the existence of the chemical bond. 
(e) arbitrary reference: The QM can learn from data corresponding to {\em any} level of theory, and even experimental results.
The main limitation of the QM is the empirical nature inherent to any statistical learning method used for inferring solutions, 
namely that meaningful predictions for new molecules can only be made if they fall in the regime of interpolation.

We believe our results to be encouraging numerical evidence that ML models can systematically infer highly 
predictive structure-property relationships from high-quality data bases generated 
via first principles atomistic simulations or experiments. 
In this study we have demonstrated the QM's performance for a rather small subset of chemical space, 
namely for small organic molecules with only up to seven atoms (not counting hydrogens) as defined by the GDB. 
Due to its inherent first principles setup we expect the overall approach to be equally applicable 
to molecules or materials of arbitrary size, configurations, and composition --- without any major modification. 
We note, however, that in order to apply the QM to other regions in chemical space with similar accuracy
differing amounts of training data might be necessary.

We conclude that combining reliable data bases with ML promises to be an important step towards
the general goal of exploring chemical compound space for the computational bottom up design of novel and improved compounds. 

\section{Acknowledgments}
This research used resources of the Argonne Leadership Computing Facility at Argonne National Laboratory,
which is supported by the Office of Science of the U.S.~DOE under contract DE-AC02-06CH11357.
K.-R.\,M.{} acknowledges partial support by DFG (MU~987/4-2) and EU (PASCAL2).
M.\,Rupp acknowledges support by FP7 programme of the European Community (Marie Curie IEF 273039). 
This work was also supported by the World Class University Program through the National Research 
Foundation of Korea funded by the Ministry of Education, Science, and Technology, under Grant R31-10008.


\section{Supplementary Materials}

True properties and geometries for all compounds. They can also be retrieved from {\tt www.quantum-machine.org}.

\section*{Appendix A: Details on Random Coulomb Matrices}

Random Coulomb matrices define a probability distribution over the set of Coulomb matrices and account for different atoms indexing of the same molecule. The following four-steps procedure randomly draws Coulomb matrices from the distribution $p(\fatM)$: (i) Take an arbitrary valid Coulomb matrix $\fatM$ of the molecule (ii) Compute the norm of each row of this Coulomb matrix: $\boldsymbol{n} = (||M_1||,\dots,||M_{23}||)$, (iii) draw a zero-mean unit-variance  noise vector $\boldsymbol{\varepsilon}$ of same size as $\boldsymbol{n}$, and (iv) permute the rows and columns of $\fatM$ with the same permutation that sorts $\boldsymbol{n} + \boldsymbol{\varepsilon}$. An important feature of random Coulomb matrices is that the probability distributions over Coulomb matrices of two different molecules are completely disjoint. This implies that the randomized representation is not introducing any noise into the prediction problem. Invariance to atoms indexing proves to be crucial for obtaining models with high predictive accuracy. The idea of encoding known invariances through such data extension has previously been used to improve prediction accuracy on image classification and handwritten digit recognition data sets\cite{ciresan}.

\section*{Appendix B: Details on Training the Neural Network}

\begin{figure}[htbp]
\includegraphics[scale=0.8, angle=0]{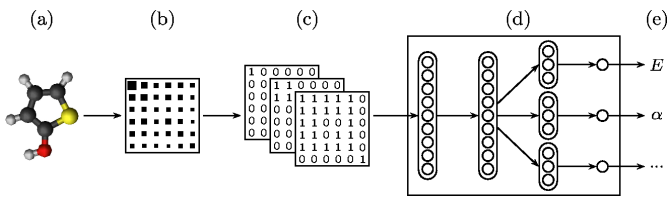}
\caption{
Predicting properties for a new molecule: 
(a) Enter Cartesian coordinates and nuclear charges, 
(b) form a Coulomb matrix,
(c) binarize representation,
(d) propagate into trained neural network, 
(e) scale outputs back to property units.
}
\label{fig:Flow}
\end{figure}

The ML model and the neural network perform a sequence of transformation to the input that are illustrated in Fig.~\ref{fig:Flow}. The Coulomb matrix is first converted to a binary representation before being processed by the neural network. The rationale for this binarization is that continuous quantities such as Coulomb repulsion energies encoded in the Coulomb matrix are best processed when their information content is distributed across many dimensions of low information content. Such binary expansion can be obtained by applying the transformation
\[
\phi(x) = \Big[...,\text{sigm}\Big(\frac{x - \theta}{\theta}\Big),\text{sigm}\Big(\frac{x}{\theta}\Big),\text{sigm}\Big(\frac{x + \theta}{\theta}\Big),...\Big]
\]
where $\phi: \mathbb{R} \rightarrow [0,1]^\infty$, the parameter $\theta$ controls the granularity of the transformation and $\text{sigm}(x) = e^x / (1+e^x)$ is a sigmoid function. Transforming Coulomb matrices $\fatM$ of size $23 \times 23$ with a granularity $\theta=1$ yields 3D tensors of size $[\infty \times 23 \times 23]$ of quasi-binary values, approximately 2000 dimensions of which are non-constant. Transforming vectors $P$ of $14$ properties with a granularity $0.25$ of same units as in Table~\ref{tab:error} yields matrices of size $[\infty \times 14]$, approximately $1000$ components of which are non-constant.

We construct a four layer neural network with $2000$, $800$, $800$ and $1000$ nodes at each layer. The network implements the function $\phi^{-1} \circ f_3 \circ f_2 \circ f_1 \circ \phi(\fatM)$ where functions $f_1$, $f_2$ and $f_3$ between each layer correspond to a linear transformation learned from data followed by a sigmoid nonlinearity. The neural network is trained to minimize the mean absolute error of each property using the stochastic gradient descent algorithm (SGD \cite{SGD}). Errors are back-propagated\cite{Backprop} from the top layer back to the inputs in order to update all parameters of the model. We run 250,000 iterations of the SGD and present at each iteration 25 training samples. During training, each molecule-property pair is presented in total $1250$ times to the neural network, but each time with a different atoms indexing. A moving average of the model parameters is maintained throughout training in order to attenuate the noise of the stochastic learning algorithm~\cite{Polyak92}. 
The moving average is set to remember the last 10\% of the training history and is used for prediction of out-of-sample molecules. 
Once the neural network has been trained, the typical CPU time for predicting all 14 properties of a new out-of-sample molecule is $\sim$100 milli seconds.
Training the neural network on a CPU takes~$\sim$24 hours. 
Prediction of an out-of-sample molecule is obtained by propagating 10 different realizations of $p(\fatM)$ and averaging outputs.
Prediction of multiple molecules can be easily parallelized by replicating the trained neural network on multiple machines.

\bibliographystyle{unsrt}
\bibliography{literatur}
\end{document}